\documentstyle[12pt]{article}
\catcode`\@=11

\@addtoreset{equation}{section}
\def\theequation{\arabic{equation}}
\def\theequation{\thesection\arabic{equation}}

\def\NPB#1#2#3{{\it Nucl.~Phys.} {\bf{B#1}} (19#2) #3}
\def\PLB#1#2#3{{\it Phys.~Lett.} {\bf{B#1}} (19#2) #3}
\def\PRD#1#2#3{{\it Phys.~Rev.} {\bf{D#1}} (19#2) #3}

\def\JHEP#1#2#3{{\it J. High Energy Phys.} {\bf#1} (19#2) #3}
\def\@normalsize{\@setsize\normalsize{15pt}\xiipt\@xiipt
\abovedisplayskip 14pt plus3pt minus3pt%
\belowdisplayskip \abovedisplayskip
\abovedisplayshortskip  \z@ plus3pt%
\belowdisplayshortskip  7pt plus3.5pt minus0pt}
\def\small{\@setsize\small{13.6pt}\xipt\@xipt
\abovedisplayskip 13pt plus3pt minus3pt%
\belowdisplayskip \abovedisplayskip
\abovedisplayshortskip  \z@ plus3pt%
\belowdisplayshortskip  7pt plus3.5pt minus0pt
\def\@listi{\parsep 4.5pt plus 2pt minus 1pt
            \itemsep \parsep
            \topsep 9pt plus 3pt minus 3pt}}

\def\underline#1{\relax\ifmmode\@@underline#1\else
        $\@@underline{\hbox{#1}}$\relax\fi}
\@twosidetrue
\relax

\catcode`@=12

\catcode`\@=11

\def\section{\@startsection{section}{1}{\z@}{3.5ex plus 1ex minus
   .2ex}{2.3ex plus .2ex}{\large\bf}}
\def\thesection{\arabic{section}.}

\def\ps@headings{\def\@oddfoot{}\def\@evenfoot{}
\def\@oddhead{\hbox{}\hfill
        \makebox[.5\textwidth]{\raggedright\ignorespaces --\thepage{}--
        \hfill }}
\def\@evenhead{\@oddhead}
\def\subsectionmark##1{\markboth{##1}{}} }

\ps@headings

\catcode`\@=12

\relax

\def\figcap{\section*{Figure Captions\markboth
        {FIGURECAPTIONS}{FIGURECAPTIONS}}\list
        {Fig. \arabic{enumi}:\hfill}{\settowidth\labelwidth{Fig. 999:}
        \leftmargin\labelwidth
        \advance\leftmargin\labelsep\usecounter{enumi}}}
 \relax
\def\tablecap{\section*{Table Captions\markboth
        {TABLECAPTIONS}{TABLECAPTIONS}}\list
        {Table \arabic{enumi}:\hfill}{\settowidth\labelwidth{Table 999:}
        \leftmargin\labelwidth
        \advance\leftmargin\labelsep\usecounter{enumi}}}
 \relax
\def\reflist{\section*{References\markboth
        {REFLIST}{REFLIST}}\list
        {[\arabic{enumi}]\hfill}{\settowidth\labelwidth{[999]}
        \leftmargin\labelwidth
        \advance\leftmargin\labelsep\usecounter{enumi}}}
 \relax

\catcode`\@=11

\def\marginnote#1{}
\newcount\hour
\newcount\minute
\newtoks\amorpm
\hour=\time\divide\hour by60
\minute=\time{\multiply\hour by60 \global\advance\minute by-
\hour}
\edef\standardtime{{\ifnum\hour<12 \global\amorpm={am}%
    \else\global\amorpm={pm}\advance\hour by-12 \fi
    \ifnum\hour=0 \hour=12 \fi
    \number\hour:\ifnum\minute<100\fi\number\minute\the\amorpm}}
\edef\militarytime{\number\hour:\ifnum\minute<100\fi\number\minute}
\def\draftlabel#1{{\@bsphack\if@filesw {\let\thepage\relax
  \xdef\@gtempa{\write\@auxout{\string
    \newlabel{#1}{{\@currentlabel}{\thepage}}}}}\@gtempa
    \if@nobreak \ifvmode\nobreak\fi\fi\fi\@esphack}
     \gdef\@eqnlabel{#1}}
\def\@eqnlabel{}
\def\@vacuum{}
\def\draftmarginnote#1{\marginpar{\raggedright\scriptsize\tt#1}}
\def\draft{\oddsidemargin -.5truein
        \def\@oddfoot{\sl preliminary draft \hfil
        \rm\thepage\hfil\sl\today\quad\militarytime}
        \let\@evenfoot\@oddfoot \overfullrule 3pt
        \let\label=\draftlabel
        \let\marginnote=\draftmarginnote
   
\def\@eqnnum{(\theequation)\rlap{\kern\marginparsep\tt\@eqnlabel}%
\global\let\@eqnlabel\@vacuum}  }
\def\preprint{\twocolumn\sloppy\flushbottom\parindent 1em
        \leftmargini 2em\leftmarginv .5em\leftmarginvi .5em
        \oddsidemargin -.5in    \evensidemargin -.5in
        \columnsep 15mm \footheight 0pt
        \textwidth 250mmin      \topmargin  -.4in
        \headheight 12pt \topskip .4in
        \textheight 175mm
        \footskip 0pt
        
\def\@oddhead{\thepage\hfil\addtocounter{page}{1}\thepage}
        \let\@evenhead\@oddhead \def\@oddfoot{} \def\@evenfoot{}  }
\def\titlepage{\@restonecolfalse\if@twocolumn\@restonecoltrue\onecolumn
     \else \newpage \fi \thispagestyle{empty}\c@page\z@
        \def\thefootnote{\fnsymbol{footnote}} }
\def\endtitlepage{\if@restonecol\twocolumn \else  \fi
        \def\thefootnote{\arabic{footnote}}
        \setcounter{footnote}{0}}  
\catcode`@=12
\relax

\def\ps@headings{\def\@oddfoot{}\def\@evenfoot{}
\def\@oddhead{\hbox{}\hfill
        \makebox[.5\textwidth]{\raggedright\ignorespaces --\thepage{}--
        \hfill }}
\def\@evenhead{\@oddhead}
\def\subsectionmark##1{\markboth{##1}{}} }

\ps@headings

\relax

\def\firstpage#1#2#3#4#5#6{
\begin{document}
\begin{titlepage}
\nopagebreak
\title{\begin{flushright}
        \vspace*{-1.8in}
        {\normalsize CERN-TH/98-382}\\[-10mm]
        {\normalsize CPTH-S691.1198}\\[-10mm]
        {\normalsize LPTHE-ORSAY 98/70}\\[-10mm]
        {\normalsize ROM2F-98/41}\\[-10mm]
        {\normalsize hep-th/9812118}\\[-4mm]
\end{flushright}
\vfill {#3}}
\author{\large #4 \\[1.0cm] #5}
\maketitle
\vskip -9mm     
\nopagebreak 
\begin{abstract} {\noindent #6}
\end{abstract}
\vfill
\begin{flushleft}
\rule{16.1cm}{0.2mm}\\[-4mm]
$^{\star}${\small Research supported in part by the EEC under TMR contract 
ERBFMRX-CT96-0090.}\\[-4mm] 
$^{\dagger}${\small Laboratoire associ{\'e} au CNRS-URA-D0063.}\\ 
CERN-TH/98-382\\ December 1998
\end{flushleft}
\thispagestyle{empty}
\end{titlepage}}
\def\simlt{\stackrel{<}{{}_\sim}}
\def\simgt{\stackrel{>}{{}_\sim}}
\newcommand{\dal}{\raisebox{0.085cm} {\fbox{\rule{0cm}{0.07cm}\,}}}
\newcommand{\dt}{\partial_{\langle T\rangle}}
\newcommand{\dtbar}{\partial_{\langle\overline{T}\rangle}}
\newcommand{\al}{\alpha^{\prime}}
\newcommand{\mst}{M_{\scriptscriptstyle \!S}}
\newcommand{\mpl}{M_{\scriptscriptstyle \!P}}
\newcommand{\dv}{\int{\rm d}^4x\sqrt{g}}
\newcommand{\lv}{\left\langle}
\newcommand{\rv}{\right\rangle}
\newcommand{\ph}{\varphi}
\newcommand{\abar}{\overline{a}}
\newcommand{\sbar}{\,\overline{\! S}}
\newcommand{\xbar}{\,\overline{\! X}}
\newcommand{\fbar}{\,\overline{\! F}}
\newcommand{\zbar}{\overline{z}}
\newcommand{\dbar}{\,\overline{\!\partial}}
\newcommand{\tbar}{\overline{T}}
\newcommand{\taubar}{\overline{\tau}}
\newcommand{\ubar}{\overline{U}}
\newcommand{\ybar}{\overline{Y}}
\newcommand{\phb}{\overline{\varphi}}
\newcommand{\cm}{Commun.\ Math.\ Phys.~}
\newcommand{\prl}{Phys.\ Rev.\ Lett.~}
\newcommand{\pr}{Phys.\ Rev.\ D~}
\newcommand{\pl}{Phys.\ Lett.\ B~}
\newcommand{\ibar}{\overline{\imath}}
\newcommand{\jbar}{\overline{\jmath}}
\newcommand{\np}{Nucl.\ Phys.\ B~}
\newcommand{\F}{{\cal F}}
\renewcommand{\L}{{\cal L}}
\newcommand{\A}{{\cal A}}
\newcommand{\e}{{\rm e}}
\newcommand{\be}{\begin{equation}}
\newcommand{\ee}{\end{equation}}
\newcommand{\ba}{\begin{eqnarray}}
\newcommand{\ea}{\end{eqnarray}}
\newcommand{\dslash}{{\not\!\partial}}
\newcommand{\gsi}{\,\raisebox{-0.13cm}{$\stackrel{\textstyle >}{\textstyle\sim}$}\,}
\newcommand{\lsi}{\,\raisebox{-0.13cm}{$\stackrel{\textstyle <}{\textstyle\sim}$}\,}
\date{}
\firstpage{3118}{IC/95/34} {\large\bf Partial breaking of 
supersymmetry, open strings and M-theory}  
{I. Antoniadis$^{\,a}$, G. D'Appollonio$^{\,b}$, E. Dudas$^{\,c,d}$ and 
A. Sagnotti$^{\,e}$} 
{\small\sl
$^a$ Centre de Physique Th{\'e}orique (CNRS UMR 7644),  Ecole Polytechnique, 
{}F-91128 Palaiseau\\[-4mm]
\small\sl$^{b}$ Dipartimento di Fisica, Univ. dell'Aquila, INFN-LNGS,
67010 Coppito,
Italy\\[-4mm]
\small\sl$^c$ TH-Division, CERN, CH-1211 Geneva 23, Switzerland\\[-3mm]
\small\sl $^d$  LPTHE$^\dagger$, B{\^a}t. 211, Univ. Paris-Sud, F-91405 Orsay\\[-4mm] 
\small\sl$^{e}$ Dipartimento di Fisica, Universit\`a di Roma ``Tor Vergata''\\[-3mm]
\small\sl INFN,
 Sezione di Roma ``Tor Vergata''\\[-4mm]
\small\sl Via della Ricerca Scientifica 1, 00133 Roma, Italy} 
{We study total and partial supersymmetry breaking by freely acting
orbifolds, or equivalently by Scherk-Schwarz  compactifications, in type
I string theory. In particular, we describe a four-dimensional chiral
compactification with spontaneously broken $N=1$ supersymmetry, some 
models with partial $N=4\to N=2$ and $N=4\to N=1$ supersymmetry
breaking and their heterotic and M-theory duals. A generic
feature of these models is that in the gravitational sector and in the
spectrum of D-branes parallel to the breaking coordinate, all mass
splittings are proportional to the compactification scale, while
global (extended) supersymmetry  remains  unbroken at tree level for the massless
excitations of D-branes transverse to the breaking direction.}
\section{Introduction}

In a recent paper \cite{ADS} we started a systematic investigation
of type I string vacua where supersymmetry is
spontaneously broken by the compactification \footnote{Previous 
work on supersymmetry breaking in open strings
is described in \cite{Ba}.}. Aside from their clear
potential interest for phenomenology, the resulting models are
additional examples of the link between closed and open
models \cite{S}-\cite{H}.
They may be constructed in a systematic fashion associating suitable
D-brane (open string) sectors to projected bulk (closed string) models
where supersymmetry breaking
is induced by a Scherk-Schwarz (SS) deformation \cite{SS}-\cite{KK}. 
In Field Theory, the SS mechanism
is realized if all fields are periodic along a compactified direction
only up to discrete R-symmetry transformations. In String Theory, the resulting
constructions are equivalent to the freely acting orbifolds discussed
in \cite{VW}, where the discrete symmetry is combined with a 
shift $\delta$ along the compact direction.

In the simplest case of toroidal compactification, the shift $\delta$
is a fraction of a lattice vector determined by the order of the
discrete symmetry ({\it e.g.} $N\delta =1$ for $Z_N$). For closed strings, 
there are two distinct choices for the shift, depending on
whether it acts identically or oppositely on left and right movers,
that are related by T-duality transformations ($R\to 1/R$ for a circle of 
radius $R$)\footnote{We consider only the
case of light-like Narain vectors, that lead to symmetry
restoration in the large (or small) radius limit.}. The SS deformation
is then recovered multiplying (or dividing) the radius $R$ by $N$, and amounts to
shifting the momenta (or the windings) of all states according to their
R-charges. In the following we will refer to these two cases as
momentum and winding shifts. The two choices
lead to very different open string spectra \cite{ADS}. Consider, for instance, a
D-brane parallel to the direction used for supersymmetry breaking. With
momentum shifts, all mass splittings are proportional to $1/R$, with $R$
the compactification radius, and to the discrete charges,
as in the closed string sector. Supersymmetry
is then restored only in the large radius limit. On the other hand, with
winding shifts all mass splittings vanish for the massless
excitations of the D-brane, while supersymmetry is restored in the
small radius limit. The physical reason for this result can be better appreciated 
in the T-dual (type I') picture. In this case, the D-brane is
orthogonal to the direction of supersymmetry breaking, that again disappears
in the decompactification limit. Therefore, it is naturally 
insensitive to the breaking of supersymmetry, as long as
one does not consider the modes, much heavier than the string scale, that 
wrap around the compact space.  On the other hand, when these modes are
excited, the D-brane does feel the breaking, since it
effectively extends into the compact space.

This phenomenon is apparently a generic feature of type I
compactifications. Therefore, it is instructive to relate these models to
heterotic compactifications using string dualities \cite{PW,HW}. 
This is possible since adiabatic arguments \cite{VW} justify the
use of duality transformations for freely acting orbifolds. 
Momentum shifts in D9 branes may then be related to ordinary
SS compactifications of the weakly coupled $SO(32)$ heterotic string.
On the other hand, winding shifts may be related to SS
deformations involving the eleventh dimension of M-theory, a non
perturbative phenomenon in the $E_8\times E_8$ heterotic
string  \cite{AQ,DG}.

In this work we explore further this line of investigation. We thus
present a first instance of a four dimensional (4d) $N=1$
chiral model with D9 and D5 branes, where supersymmetry breaking is induced by a SS
deformation based on a $Z_2$ symmetry, the spacetime fermion
parity $(-1)^F$. Our starting point is the six-dimensional supersymmetric model 
of \cite{BS,GP}, deformed as in \cite{ADS}, but here we couple
the spacetime fermion parity to the winding modes, in order
to make the connection with M-theory more transparent.
With this choice, the massless states related to the D9 brane are not affected at
tree level.

In a T-dual type-I' picture, where the D9 and D5 branes have turned into D8
and D6 branes, supersymmetry is restored in the large radius limit. Actually, this
mechanism can be realized only if the 8-branes are equally distributed
among the two fixed points of the corresponding orientifold, so that the
D8 gauge group becomes a direct product of two identical (chiral)
factors with no direct communication at the massless level. This is
similar to the type I' situation in 9 dimensions, where one can take
the large radius limit at weak string coupling only if
the branes are evenly distributed, so that the gauge group is $SO(16)\times
SO(16)$  \cite{PW}. This limit then corresponds to the decompactification of the
11th dimension of M-theory, once the two sets of D8 branes are associated
with the two 10d walls containing the two $E_8$ factors  \cite{HW}. In the
type I string, the problem manifests itself as a tadpole 
generated by the collapse of superheavy winding states that
propagate in the transverse channel and
become massless in the small radius limit. It is remarkable that these tadpole
cancellations single out uniquely the M-theory setup \cite{ADS}.

A peculiarity of our chiral models with supersymmetry breaking induced by $(-1)^F$
is a relative chirality flip, after the SS deformation, between pairs of
gauge group factors for branes perpendicular to the breaking
direction and located at different fixed points of the orientifold. 
This phenomenon was already observed in 6 dimensions \cite{ADS}, 
and is clearly induced by the modified projection of the closed sector.
However, we do not have a simple geometrical picture for it, and we are not sure 
whether it occurs in all similar constructions.

As a second step, we study the effects of SS deformations induced by
ordinary R-symmetries, rather than by the fermion parity $(-1)^F$. These
R-symmetries are discrete remnants of internal rotations in the 6d compact space
and, as explained above, the resulting deformations are equivalent to freely acting
orbifolds where internal rotations are combined with shifts \cite{KK}. Unlike the case
of $(-1)^F$, R-symmetries can break only part of the supersymmetries, 
typically one half of them,
thus yielding examples with partial supersymmetry breaking. In this
paper we work out some explicit examples of $N=4$ type I string
compactifications with partial supersymmetry breaking to $N=2$ or
$N=1$, using $Z_2$ shifts. Again, we find two distinct
options, corresponding to momentum or winding shifts, with
D-branes parallel or orthogonal to
the breaking direction, and extended supersymmetry
partly broken or unbroken for the massless excitations.

This paper is organized as follows. In Section 2 we review briefly
the 6d chiral model of ref. \cite{ADS} and present an $N=1$ reduction to
4 dimensions, obtained by a $Z_6$ orbifold compactification with SS
supersymmetry breaking induced by $(-1)^F$. In
Section 3 we present the simplest example of partial supersymmetry
breaking in 5 dimensions, using a $Z_2$ freely acting orbifold that twists 
four internal coordinates and shifts the fifth one. We first review
the parent type II string model, that breaks partially $N=8$ to
$N=4$, and then derive its open descendants, where the partial
breaking is from $N=4$ to $N=2$, in the two cases of momentum (Section 4)
and winding (Section 5) shifts.\footnote{Here we are counting the number of
supersymmetries in terms of 4d Weyl spinors.} In Section 6 we
describe a 4d type I model where $N=4$
supersymmetry is partially broken to $N=1$ using a $Z_2\times Z_2$
freely acting orbifold with three different momentum shifts. Finally,
Section 7 contains our conclusions. Throughout the paper we use the same 
conventions as in \cite{ADS}. Thus, $\alpha^\prime=2$, and for later use we 
also define (with $\eta$ the Dedekind function)
\be
Z_{m+a}(\tau) = \frac{q^{{1\over 2} {\left(\frac{m+a}{R}\right)}^2}}{\eta(\tau)}
 \ ,
\qquad
\tilde{Z}_{n+ b}(\tau) =\frac{q^{{1\over 2} {\left( (n + b){R\over 2}
\right)}^2}}{\eta(\tau)}  \label{i1}
\ee
while, in relating the direct and transverse channels, we shall use repeatedly 
the Poisson transformation
\be
\sum_m e^{2i\pi m b} Z_{m+a} (-{1 \over \tau})=R\ e^{-2i\pi ab} \sum_n e^{-2i
\pi na} {\tilde Z}_{2n+2b} ({\tau}) \ . \label{i2}
\ee
\section{A chiral four-dimensional model with $N=1 \rightarrow N=0$ breaking }
 
In this Section, using the results of \cite{ADS}, we construct a chiral 4d model with
spontaneously broken $N=1$ supersymmetry, where a portion of the massless spectrum,
that after a suitable T-duality may be associated to D9-branes, is still (globally)
supersymmetric. The starting point is the 6d $N=1$
$T^4/Z_2$ type I orbifold of \cite{BS,GP} with all five-branes at the same
fixed point, a single 6d tensor multiplet, and a gauge group $U(16)_9
\times U(16)_5$. The open string sector then contains
vector multiplets and hypermultiplets in the representations 
$({\bf 120}+{\overline{\bf 120}},{\bf1})$, $({\bf1},{\bf 120}+{\overline{\bf 120}})$ 
of the gauge group from the $99$
and $55$ sectors, and one hypermultiplet in the representation 
$({\bf 16}, {\bf 16})$ from the $59$ sector. Turning on a Wilson line, one can break
the D9 gauge group $U(16)_9$ to $U(8)_9 \times U(8)_{9'}$. Then, after a T-duality
turning D9 branes into D8 branes and D5 branes into D6 branes, one can interpret the
resulting model as M-theory compactified on $T^4 \times S^1/(Z_2 \times Z_2)$. 
In addition to the $N=2$ vector multiplets, corresponding to
the representations $({\bf 64,1,1})_R
+({\bf 1,64,1})_R+({\bf 1,1,256})_R$ of the gauge group $U(8)_8 \times U(8)_{8'}
\times U(16)_6$, the resulting massless spectrum contains the hypermultiplets
\ba
&66& : ({\bf 1},{\bf 1},{\bf 120})_L+ ({\bf 1},{\bf 1},{\overline {\bf 120}})_L \ \ ,  
\nonumber \\
&88,8'8'& : ({\bf 28},{\bf 1},{\bf 1})_L + ({\overline 
{\bf 28}},{\bf 1},{\bf 1})_L + ({\bf 1},{\bf 28},{\bf 1})_L + 
({\bf 1},{\overline {\bf 28}},{\bf 1})_L \ , \nonumber \\
&86,8'6& : ({\bf 8},{\bf 1},{\bf 16})_L + 
({\bf 1},{\bf 8},{\bf 16})_L \ . 
\label{c30}
\ea
Actually, all ND (Neumann-Dirichlet) representations occur as half-multiplets with
pairs of conjugate assignments, equivalent after charge conjugation to conventional
hypermultiplets. Thus, for instance,
the hypermultiplet assigned to the 
representation $({\bf 8},{\bf 1},{\bf 16})_L$ actually appears in the open spectrum
as a pair of half-hypermultiplets associated to the conjugate pair
$({\bf 8},{\bf 1},{\bf 16})_L+ ({{\bf {\bar 8}},{\bf 1},
{\overline {\bf 16}}})_L$. 

As shown in \cite{ADS}, starting from the supersymmetric $U(16)_9
\times U(16)_5$ model, one can break 6d supersymmetry completely in a soft fashion 
by a Scherk-Schwarz deformation along the direction that was T-dualized
above (parallel to the D6 branes and transverse to the D8 branes).  
This is induced by the
operator $(-1)^F$ acting on the lattice states, or equivalently by a
$2\pi$-rotation in a plane defined by one compact and one non-compact
coordinate. Before T-duality, the winding modes are shifted by this operation, 
while the momentum modes are unaffected.
The corresponding world-sheet current anticommutes with the 
orbifold projection, as required for the consistency of the construction
\cite{KP,a}. As in the supersymmetric case, the factorization properties of the
amplitudes and the tadpole conditions uniquely determine the gauge group
$U(8)_8 \times U(8)_{8'} \times U(16)_6$, identical to that of the
supersymmetric model. The two D8-brane factors live at
two different fixed points of the orientifold, 
and their world-volumes are orthogonal to the 
compact coordinate responsible for the breaking. The
corresponding massless matter content is
\ba
&66& : \left[ ({\bf 1,1,256})+({\bf 1},{\bf 1},{\bf 120})+ 
({\bf 1},{\bf 1},{\overline {\bf 120}}) \right]_{spin \ 0} \ , 
\nonumber \\
&88& : ({\bf 64},{\bf 1},{\bf 1})_R + ({\bf 28},{\bf 1},{\bf 1})_L + ({\overline 
{\bf 28}},{\bf 1},{\bf 1})_L \ , \nonumber \\
&8'8'& :  ({\bf 1},{\bf 64},{\bf 1})_L + ({\bf 1},{\bf 28},{\bf 1})_R + 
({\bf 1},{\overline {\bf 28}},{\bf 1})_R \ , \nonumber \\
&86,8'6& : ({\bf 8},{\bf 1},{\bf 16})_L + ({\bf 1},{{\bf 8}},{{\bf 16}})_R \ , 
\label{c0}
\ea
where the massless $66$ states are only scalars and 
the subscripts indicate the 6d chirality of the fermions. Notice that the D6 branes,
parallel to the direction of supersymmetry breaking, feel its effects at tree level, while
all massless sectors related to the D8 and D8' branes are supersymmetric.
It should be appreciated that the 6d
chirality of the fermion sectors related to the D8' branes ($8'8'$ and $8'6$ sectors)
is flipped with respect to the supersymmetric case, while all fermions related to
the D8 branes ($88$ and $86$ sectors) have the same chirality 
as in the supersymmetric case. This model could also be regarded as a compactification of
M-theory on $S^1 \times T^4/(Z_2\times Z_2)$, where each $U(8)$
factor would originate from the breaking of 
an $E_8$ gauge group, while $x_9(\equiv X)$ would be interpreted
as the eleventh dimension \cite{ADS}. However,
in addition to the gravitational sector, the $x_9$-bulk contains the D6 brane sector 
with a gauge group $U(16)$. From the M-theory viewpoint, this (bulk) gauge 
group has a nonperturbative origin, related to the
M5 brane wrapped around the compact space. 

Consider now a similar Scherk-Schwarz deformation of the 
supersymmetric 4d $N=1$ type I vacuum obtained from the type-IIB string 
compactified on $Z_6' \equiv Z_2 \times Z_3$, where
the orbifold action on the three complex internal planes is generated by
\be
Z_2 = (1 \ , \ -1 \ , \ -1 ) \ , \ Z_3 = (e^{2\pi i/3} \ , \ e^{-2\pi i/3} \ , \ 1 ) \ . 
\label{c1} 
\ee
We can introduce a $(-1)^F$ deformation acting as a $2\pi$ rotation on one 
of the coordinates
of the third complex plane and on one of the non-compact coordinates. 
This deformation is
compatible with both orbifold operations and induces the spontaneous breaking
of supersymmetry from $N=1$ to $N=0$. 
Since the $Z_3$ orbifold does not give rise to additional branes,
the open string spectrum may be simply deduced by a $Z_3$ projection of the spectrum
in eq. (\ref{c0}). 

In order to gain a better understanding of the breaking pattern, we begin by
recalling some features of the supersymmetric $Z'_6$ model. This contains $32$ D9 branes 
and $32$ D5 branes, that we can initially place at the origin \cite{Z6}.
The $Z_2$ orbifold inverts half of the Chan-Paton charges both for D9 and
for D5 branes. Thus, for instance, 
the N charges associated to D9-branes are projected according to
\be
n + \bar{n} \rightarrow i ( n - \bar{n} ) \quad .
\ee
Combining the $Z_2$ orbifold with the $Z_3$ projection
\be
n \rightarrow n_0 + e^{\frac{2 \pi i}{3}} n_1 + 
e^{-\frac{2 \pi i}{3}} n_2 \quad ,
\label{c2}
\ee
then leads to a model with N=1 supersymmetry, a 
gauge group $[U(4) \times U(4) \times U(8)]_9 \times
[U(4) \times U(4) \times U(8)]_5$, and chiral matter in the representations
\ba
99\ {\rm or} \ 55 &:& 
({\bf 4},{\bf 4},{\bf 1})+({\bar {\bf 4}},{\bar {\bf 4}},{\bf 1})+({\bar 
{\bf 4}},{\bf 4},{\bf 1})+({\bf 6},{\bf 1},{\bf 1})+
({\bf 1},{\bar {\bf 6}},{\bf 1})+
\nonumber \\ 
\ \ \ \ \ \ \ \ \  
&&({\bf 1},{\bf 1},{\bf 28})+ ({\bf 1},{\bf 1},{\overline {\bf 28}})+ 
({\bf 1},{\bf 4},{\bf 8})+({\bar {\bf 4}},{\bf 1},{\bar {\bf 8}})+
({\bf 4},{\bf 1},{\bar {\bf 8}}) + ({\bf 1},{\bar {\bf 4}},{\bf 8}) \ , \nonumber \\
59 &:& 
({\bf 1},{\bf 4},{\bf 1};{\bf 1},{\bf 4},{\bf 1})+({\bf 4},{\bf 1},{\bf 1};{\bf 1},
{\bf 1},{\bf 8})+({\bf 1},{\bf 1},{\bf 8};{\bf 4},{\bf 1},{\bf 1}) + \nonumber \\
&&({\bar {\bf 4}},{\bf 1},{\bf 1};{\bar {\bf 4}},{\bf 1},{\bf 1})+({\bf 1},{\bar{\bf 4}},
{\bf 1};{\bf 1},{\bf 1},{\bar {\bf 8}})+ 
({\bf 1},{\bf 1},{\bar {\bf 8}};{\bf 1},{\bar {\bf 4}},{\bf 1}) \ . \label{c3} 
\ea
In analogy with the 6d model, we can relate this spectrum to an M-theory 
compactification on $S^1\times T^6/(Z_2\times Z'_6)$, turning on the same Wilson line 
breaking of the D9 gauge group $U(4) \times U(4) \times U(8)$ to
$[U(2) \times U(2)\times U(4)]^2$ and then performing a T-duality to the type I'
description.
The resulting open spectrum can be obtained by the $Z_3$ truncation (\ref{c2})
of the 6d spectrum (\ref{c30}), and thus contains the following $N=1$ massless chiral 
matter multiplets (in a self-explanatory shorthand notation):
\ba
66 &:&  ({\bf 4},{\bf 4},{\bf 1})+({\bar {\bf 4}},{\bar {\bf 4}},{\bf 1})+({\bar 
{\bf 4}},{\bf 4},{\bf 1})+({\bf 6},{\bf 1},{\bf 1})+
({\bf 1},{\bar {\bf 6}},{\bf 1})+ \nonumber \\ 
\ \ \ \ \ \ \ \ \  
&&({\bf 1},{\bf 1},{\bf 28})+ ({\bf 1},{\bf 1},{\overline {\bf 28}})+ 
({\bf 1},{\bf 4},{\bf 8})+({\bar {\bf 4}},{\bf 1},{\bar {\bf 8}})+
({\bf 4},{\bf 1},{\bar {\bf 8}}) + ({\bf 1},{\bar {\bf 4}},{\bf 8}) \ , \nonumber \\
88 \ {\rm or} \ 8'8' &:& ({\bar {\bf 2}},{\bf 2},{\bf 1})+({\bf 2},{\bf 2},{\bf 1})+
({\bar {\bf 2}},{\bar 
{\bf 2}},{\bf 1})+({\bf 2},{\bf 1},{\bar {\bf 4}})+({\bar {\bf 2}},{\bf 1},{\bar {\bf 4}})
+ \nonumber \\
&&({\bf 1},{\bar {\bf 2}},{\bf 4})+({\bf 1},{\bf 2},{\bf 4})+({\bf 1},{\bf 1},{\bf 6})+
({\bf 1},{\bf 1},{\bar {\bf 6}})+({\bf 1},{\bf 1},{\bf 1})+({\bf 1},{\bar {\bf 1}},{\bf 1})
 \ , \nonumber \\
86 \ {\rm or} \ 8'6 &:&({\bf 1},{\bf 2},{\bf 1};{\bf 1},{\bf 4},{\bf 1})+({\bf 1},
{\bf 1},{\bf 4};{\bf 4},{\bf 1},{\bf 1})+({\bf 2},{\bf 1},{\bf 1};{\bf 1},{\bf 1},{\bf 8}) 
+ \nonumber \\
&& ({\bar {\bf 2}},{\bf 1},{\bf 1};{\bar {\bf 4}},{\bf 1},{\bf 1})+
({\bf 1},{\bf 1},{\bar {\bf 4}};{\bf 1},{\bar {\bf 4}},{\bf 1})+
({\bf 1},{\bar {\bf 2}},{\bf 1};{\bf 1},{\bf 1},{\bar {\bf 8}}) \ , \label{c4}
\ea 
where by convention,
all the 4d fermions displayed above are left-handed, and ``barred" 
representations have $U(1)$ charges opposite to those of ``unbarred" ones.

The open spectrum of the deformed, non-supersymmetric $Z_6'$ model, can similarly 
be obtained from the 6d spectrum of eq. (\ref{c0}) by the $Z_3$ truncation (\ref{c2}).            
In M-theory language, where $x_9(\equiv X)$ would be interpreted
as the eleventh dimension, the identical D8 and D8' groups 
$[U(2) \times U(2) \times U(4)]$, of the Pati-Salam type,  
would result from the breaking of the two $E_8$ factors. However, 
as in the supersymmetric case, 
the $x_9$-bulk contains both the gravitational sector and the D6 brane sector, 
with a gauge
group $U(4) \times U(4) \times U(8)$. From the M-theory viewpoint, this gauge 
group is of non-perturbative origin, and results from the wrapping of the
M5 brane around the compact space. Since this group
lives in the bulk, it feels at tree-level the breaking of supersymmetry along the
eleventh dimension. Therefore, in the D6 brane sector 
all fermions become massive at tree level and, in addition to the 
$U(4) \times U(4) \times U(8)$ gauge bosons, the massless spectrum
contains only complex scalars in the representations 
\ba
66 &:& 
({\bf 4},{\bf 4},{\bf 1})+({\bar {\bf 4}},{\bar {\bf 4}},{\bf 1})+({\bar 
{\bf 4}},{\bf 4},{\bf 1})+({\bf 6},{\bf 1},{\bf 1})+
({\bf 1},{\bar {\bf 6}},{\bf 1})+ \nonumber \\ 
\ \ \ \ \ \ \ \ \  
&&({\bf 1},{\bf 1},{\bf 28})+ ({\bf 1},{\bf 1},{\overline {\bf 28}})+ 
({\bf 1},{\bf 4},{\bf 8})+({\bar {\bf 4}},{\bf 1},{\bar {\bf 8}})+
({\bf 4},{\bf 1},{\bar {\bf 8}}) + ({\bf 1},{\bar {\bf 4}},{\bf 8}) \ . \nonumber \\
\label{c6}
\ea

On the other hand, the massless spectra of the D8 and D8' branes, both transverse to $x_9$,
are supersymmetric. In addition to the $N=1$ gauge multiplets, the D8 brane sector contains 
the massless chiral representations 
\ba
88 &:& ({\bar {\bf 2}},{\bf 2},{\bf 1})+({\bf 2},{\bf 2},{\bf 1})+({\bar {\bf 2}},{\bar 
{\bf 2}},{\bf 1})+({\bf 2},{\bf 1},{\bar {\bf 4}})+({\bar {\bf 2}},{\bf 1},{\bar {\bf 4}})
+ \nonumber \\
&&({\bf 1},{\bar {\bf 2}},{\bf 4})+({\bf 1},{\bf 2},{\bf 4})+({\bf 1},{\bf 1},{\bf 6})+
({\bf 1},{\bf 1},{\bar {\bf 6}})+({\bf 1},{\bf 1},{\bf 1})+({\bf 1},{\bar {\bf 1}},{\bf 1})
 \, . \label{c5}
\ea 
The massless spectrum
of the D8' gauge factor contains the corresponding conjugate representations, 
a phenomenon that can be traced
to the chirality change in the parent 6d model.
There are no 88' massless representations mixing the two ``boundary'' gauge factors.

The
mixed 86 and 8'6 sectors, also supersymmetric at tree level, contain
the massless representations 
\ba
86 &:&({\bf 1},{\bf 2},{\bf 1};{\bf 1},{\bf 1},{\bf 1};
{\bf 1},{\bf 4},{\bf 1})+({\bf 1},{\bf 1},{\bf 4};{\bf 1},{\bf 1},{\bf 1};
{\bf 4},{\bf 1},{\bf 1})+({\bf 2},{\bf 1},{\bf 1};{\bf 1},{\bf 1},{\bf 1};
{\bf 1},{\bf 1},{\bf 8}) + \nonumber \\
&& ({\bar {\bf 2}},{\bf 1},{\bf 1};{\bf 1},{\bf 1},{\bf 1};
{\bar {\bf 4}},{\bf 1},{\bf 1})+
({\bf 1},{\bf 1},{\bar {\bf 4}};{\bf 1},{\bf 1},{\bf 1};
{\bf 1},{\bar {\bf 4}},{\bf 1})+
({\bf 1},{\bar {\bf 2}},{\bf 1};{\bf 1},{\bf 1},{\bf 1};
{\bf 1},{\bf 1},{\bar {\bf 8}}) \, ,\nonumber \\
8'6 &:&({\bf 1},{\bf 1},{\bf 1};{\bar{\bf 1}},{\bar{\bf 2}},{\bar{\bf 1}};
{\bar{\bf 1}},{\bar{\bf 4}},{\bar{\bf 1}})+({\bf 1},{\bf 1},{\bf 1};
{\bar{\bf 1}},{\bar{\bf 1}},{\bar{\bf 4}};
{\bar{\bf 4}},{\bar{\bf 1}},{\bar{\bf 1}})+
({\bf 1},{\bf 1},{\bf 1};{\bar{\bf 2}},{\bar{\bf 1}},{\bar{\bf 1}};
{\bar{\bf 1}},{\bar{\bf 1}},{\bar{\bf 8}}) + \nonumber \\
&& ({\bf 1},{\bf 1},{\bf 1};{\bf 2},{\bar{\bf 1}},{\bar{\bf 1}};
{\bf 4},{\bar{\bf 1}},{\bar{\bf 1}})+
({\bf 1},{\bf 1},{\bf 1};{\bar{\bf 1}},{\bar{\bf 1}},{\bf 4};
{\bar{\bf 1}},{\bf 4},{\bar{\bf 1}})+
({\bf 1},{\bf 1},{\bf 1};{\bar{\bf 1}},{\bf 2},{\bar{\bf 1}};
{\bar{\bf 1}},{\bar{\bf 1}},{\bf 8}) \, ,
\label{c7}
\ea
where the labels refer to the nine factors of
the gauge group $[U(2) \times U(2) \times U(4)]_8 \times 
[U(2) \times U(2) \times U(4)]_{8'}\times
[U(4) \times U(4) \times U(8)]_6$. Notice again the change of chirality (or complex
conjugation) in the representations of the D8' gauge group compared to the
supersymmetric case (\ref{c4}).
Of course, one can interchange the roles of D9 and D5 branes in this model 
simply
coupling (in type-I language) the supersymmetry breaking deformation to the momentum modes 
of $X$, rather than to the winding modes.

\section{$N=4 \rightarrow N=2$ supersymmetry breaking in the closed sector}

In \cite{ADS} and in the previous Section we have considered a
Scherk-Schwarz deformation along a compact coordinate $X$ induced by the 
fermion parity $(-1)^F$. As we have seen, this breaks 
supersymmetry completely in the closed sector and
in the open sectors that correspond to D-branes parallel to the 
coordinate $X$. In this Section we consider the effects of other (discrete)
symmetries, such as ordinary R-symmetries, that correspond to global rotations in the 6d
internal space. As we shall see,  in general these deformations
result in partial breakings of supersymmetry. For heterotic and type 
II strings, models of this type were first 
constructed by Kiritsis and Kounnas \cite{KK}, who also showed their equivalent 
interpretation as freely-acting orbifolds \cite{VW}.
Their results are our starting point for the construction of open
descendants. 

For pedagogical reasons, we begin our study from 5d compactifications. 
These provide the
simplest examples of partial supersymmetry breaking from $N=2$ to $N=1$ in 5d, 
or equivalently, after an additional circle reduction to four dimensions, 
from $N=4$ to $N=2$.
We thus consider the type IIB string compactified on a five-torus 
$T^4\times S^1$
and couple the momenta or the windings along the $S^1$ coordinate
$x_5$ to the charges of a discrete symmetry associated to a current $J$, that
induce a simultaneous $\pi$ rotation on the two complex internal planes
$X_3\equiv x_6+i x_7$ and $X_4\equiv x_8+ix_9$ and on their world-sheet
fermionic superpartners $\Psi_3\equiv\psi_6+i\psi_7$ and $\Psi_4\equiv\psi_8+i\psi_9$. 
Alternatively, under this rotation all four bosonic and fermionic coordinates of $T^4$, 
$x_6,\dots,x_9$ and $\psi_6,\dots,\psi_9$, change sign. Denoting by $|s_1s_2s_3s_4>$ the 
spacetime fermions, with $s_i(= \pm 1/2)$ the external $(s_1,s_2)$ and internal
$(s_3,s_4)$ helicities with respect to the $SO(2)^4$ decomposition of the 10d $SO(8)$
little group, one can represent the action of the deformation
on massless states as $exp(i\pi \oint J) = exp[i \pi (s_3+s_4)]$. 
Obviously, this does not affect
the massless gravitini  with $s_3=-s_4$, and therefore breaks only 
half of the supersymmetries. Taking into 
account the $\Omega$ projection for type I descendants,  
after an additional circle reduction to 4d this describes a partial
breaking from $N=4$ to $N=2$. 

With a momentum shift, in the zero-winding sector the SS deformation reduces to
the field-theoretic boundary conditions for higher dimensional fields $\Phi$:
\be
\Phi(x_5+2\pi R)=U\Phi(x_5)\quad ; \qquad U=e^{i\pi\oint J}\ \ \ (U^2=1)\ .
\label{bc}
\ee
In the rest of the closed spectrum, the SS deformation is then 
uniquely determined by modular invariance, and
the torus partition function reads \cite{KK} 
\be
T = T_{N=4}+ 
{\sum_{{\tilde M},N}}^\prime {\left| \sum_{a,b=0,1/2} (-1)^{2(a+b+ab)}
\frac{\theta^2\pmatrix{a\cr b} 
\theta\pmatrix{a - N/2\cr b + {\tilde M}/2} 
\theta\pmatrix{a + N/2\cr b - {\tilde M}/2}}{\eta^6 \theta\pmatrix{1/2 - 
N/2\cr 1/2 + {\tilde M}/2}
\theta\pmatrix{1/2 + N/2\cr 1/2 - {\tilde M}/2}}\right|}^2 Z_{{\tilde M},N} 
\ , \label{p1}
\ee
where $T_{N=4}$ is the supersymmetric type IIB torus partition function on $T^4\times
S^1$, and $Z_{{\tilde M},N}$, with ${\tilde M}$ 
the Poisson resummed
momentum, is the $S^1$ lattice sum for the circle coordinate
$X(\equiv x_5)$ coupled to the $J$-charges.
In addition, the $\theta$'s are Jacobi theta-functions depending on the 
modular parameter
$\tau$ of the world-sheet torus and
the ``primed'' sum over $({\tilde M},N)$ in the second term in (\ref{p1}) does not
contain the $(even,even)$ part.
After the redefinitions $N = 2n + k$ and ${\tilde M} = 2{\tilde m} + l$ with $l,k = 0,1$, 
one finds (omitting the contribution of the transverse bosons) 
\ba
&&T = \frac{1}{2} 
\{ |Q_O + Q_V|^2 \Lambda_{4,4} Z_{2{\tilde m},2n} 
+ |Q_O - Q_V|^2 |I_4I_4 - V_4V_4|_B^2 Z_{2{\tilde m}+1,2n} \nonumber \\
&& + |Q_S + Q_C|^2 |Q_S + Q_C|_B^2 Z_{2{\tilde m},2n+1} + 
|Q_S - Q_C|^2 |Q_S - Q_C|_B^2 Z_{2{\tilde m}+1,2n+1} \}  \quad ,
\label{p2}
\ea
where $\Lambda_{4,4}$ is the lattice sum for $T^4$ and,
following \cite{BS}, we have introduced the convenient notation
\ba 
Q_O = V_4I_4-C_4C_4 \ , \qquad Q_V = I_4V_4-S_4S_4 \ , \nonumber\\ 
Q_S = I_4C_4-S_4I_4 \ , \qquad Q_C = V_4S_4-C_4V_4 \ . \label{p3} 
\ea
$I_{n},V_{n},S_{n},C_{n}$ denote the standard $SO(2n)$ level-one characters, corresponding
to the conjugacy classes of the identity, the vector and the two chiral spinors,
while the subscript $B$ refers to the compact bosonic modes, fermionized according to
\ba 
4 {\eta^2 \over \theta_2^2} &=& {\theta_3^2 \theta_4^2 \over
\eta^4}=(I_4I_4-V_4V_4)_B
\ \ , \ \ 4 {\eta^2 \over \theta_4^2} = {\theta_2^2 \theta_3^2 \over \eta^4}=
(Q_S+Q_C)_B \ ,
\nonumber \\ 4 {\eta^2 \over \theta_3^2} &=& {\theta_2^2 \theta_4^2 \over \eta^4}=
(Q_S-Q_C)_B \ .
\label{p30}
\ea 

After a Poisson resummation in ${\tilde m}$, one can recast eq. (\ref{p2}) in the 
form\footnote{From now on we omit the contribution of the
transverse space-time bosons.}
\ba
T \!&=&\! \frac{1}{2} 
\{ |Q_O \!+\! Q_V|^2 \Lambda_{4,4} (Z_{m,2n} \!+\!
Z_{m+{1 \over 2},2n})   
\!+\! |Q_O \!-\! Q_V|^2 |I_4I_4 \!-\! V_4V_4|_B^2 (Z_{m,2n} \!-\!
Z_{m+{1 \over 2},2n})   \nonumber \\
&& + |Q_S + Q_C|^2 |Q_S + Q_C|_B^2 (Z_{m,2n+1} +
Z_{m+{1 \over 2},2n+1})   \nonumber \\ 
&& + |Q_S - Q_C|^2 |Q_S - Q_C|_B^2 (Z_{m,2n+1} -
Z_{m+{1 \over 2},2n+1})    \} \quad .
\label{p4}
\ea
As discussed in \cite{KK}, after halving the radius $R$ to $R/2$
this can be reinterpreted as the partition function of
a freely-acting orbifold. As can be seen from eq. (\ref{bc}), in this case
the relevant orbifold is $IIB/(-1)^m {\cal I}$, where  $(-1)^m$ denotes
the order-two shift $x_5 \rightarrow x_5 + \pi R$ and
${\cal I}$ denotes the inversion of the 
four internal coordinates, ${\cal I} x_{6,\dots,9} = - x_{6,\dots,9}$. Our notation is 
particularly convenient, since ${\cal I}$
has a very simple action on the basis (\ref{p3}): it inverts $Q_V$ and 
$Q_C$ and leaves $Q_O$ and $Q_S$ unaffected. The resulting form of the
torus amplitude,
\ba
T &=& \frac{1}{2} 
\{ |Q_O + Q_V|^2 \Lambda_{4,4} Z_{m,n} 
+ |Q_O - Q_V|^2 |I_4I_4 - V_4V_4|_B^2 (-1)^m Z_{m,n} \nonumber \\
& + & |Q_S + Q_C|^2 |Q_S + Q_C|_B^2 Z_{m,n+{1 \over 2}}+ 
|Q_S - Q_C|^2 |Q_S - Q_C|_B^2 (-1)^m Z_{m,n+{1 \over 2}} \} \, ,
\label{p5}
\ea
will be the starting point for the construction of the Scherk-Schwarz breaking
model in the next Section.

Alternatively, one can couple the $J$-charges to the winding modes of $X$. 
The corresponding torus
amplitude, simply obtained interchanging $m$ and $n$ in (\ref{p4}) and (\ref{p5}), 
will be our starting point for the construction of the M-theory breaking
model in Section 5. 
\section{$N=4 \rightarrow N=2$ Scherk-Schwarz breaking }

Using (\ref{p5}) and the standard methods of \cite{S,PS,BS}, one can obtain 
the direct-channel Klein bottle amplitude 
\be
K \!=\! \frac{1}{4}(Q_O + Q_V) [ (P \!+\! W) Z_{2m} + (P \!-\! W)Z_{2m+1} ]
  \!=\! \frac{1}{4}(Q_O + Q_V) [ P Z_{m} + W(-1)^m Z_{m} ] \ ,
\label{s0}
\ee
where $P$ and $W$ denote the momentum and winding sums for $T^4$ and $Z_m$ denotes
the momentum sum along $X$. Notice that, in the Klein 
bottle, the orbifold projection acts effectively as $(-1)^m$. 
In the three relevant 
amplitudes (Klein bottle, annulus and M\"obius-strip), the relations between 
the direct-channel parameter $t$ and the transverse-channel
parameter $l$ are
\ba
&K& \ : \ \tau = 2it_K \ , \qquad l = {1 \over 2t_K} \ , \nonumber \\
&A& \ : \ \tau = {it_A \over 2} \ , \qquad l = {2 \over t_A} \ , \nonumber \\
&M& \ : \ \tau = {it_M \over 2}+{1 \over 2} \ , \qquad l = {1 \over 2t_M} \ \ . \label{s1}
\ea
The direct and transverse Klein-bottle and annulus amplitudes are thus related
by an $S$ transformation while, in a real basis of ``hatted'' characters, 
for the M\"obius strip the corresponding transformation is effected by the
sequence $P= T^{1/2} S T^2 S T^{1/2}$  \cite{BS}. Here, as usual, $S$ denotes the 
modular inversion,
$\tau \rightarrow -1/\tau$, while $T$ denotes the modular shift $\tau \rightarrow 
\tau + 1$.

The transverse-channel Klein bottle amplitude is then
\be
\tilde{K} = \frac{2^5}{4}R(Q_O + Q_V)[v_4 W^e \tilde{Z}_{2n} +
\frac{P^e}{v_4} \tilde{Z}_{2n+1}]  \quad ,
\label{s2}
\ee
with $P^e$ ($W^e$) the even momentum (winding) sums for the $T^4$ lattice 
and $v_4$ the $T^4$ volume. Since the only massless 
tadpole contribution in $\tilde{K}$ 
is proportional to $v_4$, one can anticipate that the open spectrum will contain 
only D9 branes.
 
>From (\ref{p5}) one can also read the transverse-channel annulus amplitude. Introducing the
minimum number of independent reflection coefficients $\alpha, \beta, \gamma, \delta$, 
\ba
\tilde{A} &=& \frac{2^{-5}}{4} R \{  
[ (Q_O + Q_V) ( \alpha^2 v_4 W + \beta^2 \frac{P}{v_4} )
+ 2 \alpha \beta (Q_O - Q_V)(I_4I_4 - V_4V_4)_B  ] \tilde{Z_n} \nonumber \\
&+& [ (\gamma^2+ \delta^2)(Q_S + Q_C)(Q_S + Q_C)_B
+ 2 \gamma \delta (Q_S - Q_C)(Q_S - Q_C)_B] \tilde{Z}_{n + {1 \over 2}} \}
\ . \label{s3}
\ea
In the direct-channel amplitude corresponding to (\ref{s3}), $\alpha$ and $\gamma$
are related to Neumann (N) charges, while $\beta$ and $\delta$ are related to
Dirichlet (D) charges. Thus the cross terms, proportional to $\alpha \beta$ and
$\gamma \delta$, determine the ND sector of the model.

A basic test of the consistency of these amplitudes is obtained from the terms
at the origin of the $T^4$ lattice, whose reflection coefficients are to arrange
in perfect squares. In the following three equations, 
for notational simplicity we omit 
the $B$ subscript for the  contributions of the internal bosons. Then,
\be
\tilde{K}_0 = \frac{2^5}{4}R [(Q_OI_4I_4 + Q_VV_4V_4)
+ (Q_OV_4V_4 + Q_VI_4I_4)] 
( v_4 \tilde{Z}_{2n} + \frac{1}{v_4}\tilde{Z}_{2n+1})   \quad 
\label{s4}
\ee
and
\be
\tilde{A}_0 = \frac{2^{-5}}{4}R [ (Q_OI_4I_4 + Q_VV_4V_4)
( \sqrt{v_4} \alpha + \frac{\beta}{\sqrt{v_4}})^2 \tilde{Z}_n
+ (Q_OV_4V_4 + Q_VI_4I_4) 
( \sqrt{v_4} \alpha - \frac{\beta}{\sqrt{v_4}})^2 \tilde{Z}_n ] \quad .
\label{s5}
\ee
The geometric mean of the characters common to $\tilde{K}_0$ and
$\tilde{A}_0$ determines the transverse-channel M\"obius amplitude at 
the origin of the $T^4$ lattice,
\ba
\tilde{M}_0 &=& - \frac{R}{2} \{ 
[ \sqrt{v_4}( \sqrt{v_4} \alpha + \frac{\beta}{\sqrt{v_4}}) \tilde{Z}_{2n}
+ \frac{1}{\sqrt{v_4}} ( \sqrt{v_4} \alpha + \frac{\beta}{\sqrt{v_4}}) 
\tilde{Z}_{2n+1} ] (\hat{Q}_O\hat{I}_4\hat{I}_4 
+ \hat{Q}_V\hat{V}_4\hat{V}_4)  \nonumber \\
&+& [ \sqrt{v_4}( \sqrt{v_4} \alpha - \frac{\beta}{\sqrt{v_4}}) \tilde{Z}_{2n}
- \frac{1}{\sqrt{v_4}} ( \sqrt{v_4} \alpha - \frac{\beta}{\sqrt{v_4}}) 
\tilde{Z}_{2n+1} ] 
(\hat{Q}_O\hat{V}_4\hat{V}_4 + \hat{Q}_V\hat{I}_4\hat{I}_4) \} \, ,
\label{s6}
\ea
where in the ``hatted'' characters, defined as in \cite{ADS},
the argument is shifted by $1/2$, consistently with 
eq. (\ref{s1}) \cite{PS,BS}.
>From the contributions at the origin, one can deduce the 
full transverse-channel M\"obius amplitude:
\be
\tilde{M} \!\!=\!- \frac{R}{2} [ (\hat{Q}_O + \hat{Q}_V)( \alpha v_4 W^e \tilde{Z}_{2n}
+ \beta \frac{P^e}{v_4}\tilde{Z}_{2n+1})
+ (\hat{Q}_O - \hat{Q}_V) 
(\hat{I}_4\hat{I}_4 - \hat{V}_4\hat{V}_4)_B 
( \beta \tilde{Z}_{2n} + \alpha \tilde{Z}_{2n+1} )]
\, .
\label{s7}
\ee

Combining eqs. (\ref{s2}), (\ref{s3}) and (\ref{s7}), one can then derive the tadpole
conditions for $\alpha$ and $\beta$,
\be
\alpha = 32 \quad , \ \ \  \beta = 0 
\quad . \label{s8}
\ee
Thus $\delta$, the orbifold breaking of $\beta$, must also vanish, and
one is led to the parametrization
\be
\alpha = n_1 + n_2 = 32 \quad , \hspace{1cm} \gamma = n_1 - n_2  \quad ,
\label{s9}
\ee
in terms of which, after doubling the radius to return to the
original Scherk-Schwarz basis (\ref{p4}), the open string amplitudes 
are\footnote{With our convention $\alpha'=2$, in the direct channel 
annulus and M\"obius amplitudes $Z_{2m}$ describes integer momentum levels, 
while $Z_{2m+1}$ describes half-integer momentum levels.}
\ba
A \!\!\!\!&=&\!\!\!\! \frac{(n_1+n_2)^2}{4} (Q_O \!+\! Q_V) P Z_m +
{(n_1-n_2)^2 \over 4} (Q_O \!-\! Q_V)(I_4I_4 \!-\! V_4V_4)_B (-1)^m Z_m 
\ , \nonumber \\
M \!\!\!\!&=&\!\!\!\! - {(n_1+n_2) \over 4} \{ (\hat{Q}_O \!+\! \hat{Q}_V) P 
Z_m \!+\!
(\hat{Q}_O \!-\! \hat{Q}_V)(\hat{I}_4 \hat{I}_4 \!-\! \hat{V}_4 \hat{V}_4)_B 
(-1)^m Z_m  \} \ . \label{s10}
\ea
The resulting model has $N=2$ supersymmetry in 4d and a gauge group $SO(n_1) \times
SO(n_2)$ originating from D9 branes. This
is  manifestly a Scherk-Schwarz deformation of the $N=4$ supersymmetric 6d 
type I model
with a Wilson line that breaks the $SO(32)$ gauge group of the D9 branes
to $SO(n_1) \times SO(n_2)$, in analogy with the first model 
described in \cite{ADS}. For instance, at the origin of the $T^4$ lattice, all
integer ($2m$) momentum levels contain the adjoint vector multiplets and one 
hypermultiplet in the representation $({\bf n_1,n_2})$, while all half-integer 
momentum levels contain one vector multiplet in the $({\bf n_1,n_2})$ representation 
and hypermultiplets in the adjoint
representations $({\bf n_1(n_1-1)/2,1})+({\bf 1,n_2(n_2-1)/2})$. The particular choice $n_2=0$ corresponds to a trivial 
Wilson line, and thus to the unbroken $SO(32)$ D9 gauge group.  
As expected, $N=4$ supersymmetry is recovered in the $R \rightarrow
\infty$ limit, without any additional constraints on the models. 

\section{$N=4 \rightarrow N=2$ M-theory breaking}

In this case the starting point is
\ba
 T \!\!&=& \!\!\frac{1}{2} 
\{ |Q_O + Q_V|^2 \Lambda_{4,4} Z_{m,n} 
+ |Q_O - Q_V|^2 |I_4I_4 - V_4V_4|_B^2 (-1)^n Z_{m,n} \nonumber \\
\!\!&+& \!\!|Q_S + Q_C|^2 |Q_S + Q_C|_B^2 Z_{m+{1 \over 2},n}
+ |Q_S - Q_C|^2 |Q_S - Q_C|_B^2 (-1)^n Z_{m+{1 \over 2},n} \}  \ ,
\label{m1}
\ea
with the understanding that the Scherk-Schwarz (M-theory) basis is obtained after 
halving the radius. 
Following the same steps as in the first model, one finds the direct channel 
Klein-bottle amplitude
\be
K = \frac{1}{4}[(Q_O + Q_V)(P+W) Z_m + 2 (Q_S + Q_C)(Q_S + Q_C)_B Z_{m+{1 \over 2}}]
\quad 
\label{m2}
\ee
and, after an S-transformation, the transverse-channel amplitude
\be
\tilde{K} = \frac{2^5}{4}R [ (Q_O + Q_V) (v_4W^e + \frac{P^e}{v_4}) 
\tilde{Z}_{2n}
+ 2(Q_O - Q_V) (I_4I_4-V_4V_4)_B(-1)^n\tilde{Z}_{2n}] \quad ,
\label{m3}
\ee
where the superscript $e$ identifies the even terms in the $T^4$ lattice
sums.
>From the massless $1/v_4$ tadpole in $\tilde{K}$, one may anticipate the existence 
of one set of D5 branes, in addition to the D9 branes related to the
tadpole contribution proportional to $v_4$.

Starting from (\ref{m1}), one can then find the transverse annulus amplitude. Introducing
the minimum required number of Chan-Paton charges $(\alpha_1,\alpha_2,\beta_1,\beta_2)$
in a convenient way,
\ba
\tilde{A} \!\!&=&\!\! \frac{2^{-5}}{4}R \{ 
[(Q_O + Q_V)(\alpha_1^2Wv_4 + \beta_1^2 \frac{P}{v_4}) + 
2 \alpha_1 \beta_1(Q_O - Q_V) (I_4I_4-V_4V_4)_B] \tilde{Z}_{4n} \nonumber \\
\!\!&+&\!\! [(Q_O + Q_V)(\alpha_2^2Wv_4 + \beta_2^2 \frac{P}{v_4}) -
2 \alpha_2 \beta_2(Q_O - Q_V) (I_4I_4-V_4V_4)_B] \tilde{Z}_{4n+2} \} \, \, .
\label{m4}
\ea
As usual, the origin of the $T^4$ lattice allows one to identify the 
M\"obius amplitude. In the following three equations, we omit again
the $B$ subscript for the  contributions of the internal bosons. The geometric
mean of the coefficients of the characters common to
\ba
\tilde{K}_0 \!\!\!\! &=&\!\!\! \frac{2^5}{4}R  \{ [ (Q_OI_4I_4\!+\!Q_VV_4V_4)
(\sqrt{v_4} + \frac{1}{\sqrt{v_4}})^2 
+ (Q_OV_4V_4\!+\!Q_VI_4I_4)
(\sqrt{v_4} - \frac{1}{\sqrt{v_4}})^2 ] \tilde{Z}_{4n} \nonumber \\
\!\!\!\!\!\!&+&\!\!\! [ (Q_OI_4I_4+Q_VV_4V_4)
(\sqrt{v_4} - \frac{1}{\sqrt{v_4}})^2 
\!+\! (Q_OV_4V_4+Q_VI_4I_4)
(\sqrt{v_4} + \frac{1}{\sqrt{v_4}})^2 ] \tilde{Z}_{4n+2} \} \quad 
\label{m5}
\ea
and
\ba
 \tilde{A}_0 \!\!\!\!&=&\!\!\!\! \frac{2^{-5}}{4}R \{ [ (Q_OI_4I_4\!+\!Q_VV_4V_4)
( \alpha_1 \sqrt{v_4} \!+\!  \frac{\beta_1}{\sqrt{v_4}})^2 
\!+\! (Q_OV_4V_4\!+\!Q_VI_4I_4)
( \alpha_1 \sqrt{v_4} \!-\!  \frac{\beta_1}{\sqrt{v_4}})^2 ] \tilde{Z}_{4n} \nonumber \\
 \!\!\!\!\!&+&\!\!\!\!\! [ (Q_OI_4I_4\!+\!Q_VV_4V_4)
( \alpha_2 \sqrt{v_4} \!-\!  \frac{\beta_2}{\sqrt{v_4}})^2 \!+\! (Q_OV_4V_4\!+\!Q_VI_4I_4)
( \alpha_2 \sqrt{v_4}\!+\! \frac{\beta_2}{\sqrt{v_4}})^2 ] \tilde{Z}_{4n+2} \}
\label{m6}
\ea
determines in the transverse M\"obius amplitude the contributions                              at the origin of the
$T^4$ lattice:
\ba
\tilde{M}_0 \!\!\!\!&=&\!\!\! - \frac{R}{2} \{ (\hat{Q}_O\hat{I}_4\hat{I}_4 
\!+\! \hat{Q}_V\hat{V}_4\hat{V}_4)  
[( \alpha_1v_4 \!+\! \frac{\beta_1}{v_4} \!+\! \alpha_1 \!+\! \beta_1 ) \tilde{Z}_{4n}
\!+\! ( \alpha_2v_4 \!+\! \frac{\beta_2}{v_4} \!-\! \beta_2 \!-\! \alpha_2 ) 
\tilde{Z}_{4n+2} ] \nonumber \\
\!\!\!\!\!\!\!\!\!\!\!\!&+&\!\!\!\!(\hat{Q}_O\hat{V}_4\hat{V}_4 \!+\! \hat{Q}_V\hat{I}_4\hat{I}_4)
[  ( \alpha_1v_4 \!+ \frac{\beta_1}{v_4} \!-\! \alpha_1 \!-\! \beta_1 ) \tilde{Z}_{4n} \!
+ \!( \alpha_2v_4 + \frac{\beta_2}{v_4} + \alpha_2 + \beta_2 ) 
\tilde{Z}_{4n+2} ]  \} \, .
\label{m7}
\ea
These suffice to reconstruct the full transverse-channel M\"obius amplitude, that reads
\ba
 \tilde{M} \!\!\!\!&=&\!\!\! -\frac{R}{2} \{ 
[(\hat{Q}_O + \hat{Q}_V)(\alpha_1W^ev_4 + \beta_1 \frac{P^e}{v_4}) + 
(\alpha_1 + \beta_1) (\hat{Q}_O - \hat{Q}_V) 
(\hat{I}_4\hat{I}_4-\hat{V}_4\hat{V}_4)_B] \tilde{Z}_{4n} \nonumber \\
 \!\!\!\!&+&\!\!\! [(\hat{Q}_O + \hat{Q}_V)(\alpha_2W^ev_4 + \beta_2 \frac{P^e}{v_4}) 
 - ( \alpha_2 + \beta_2 ) (\hat{Q}_O - \hat{Q}_V) 
(\hat{I}_4\hat{I}_4-\hat{V}_4\hat{V}_4)_B] \tilde{Z}_{4n+2} \} \, .
\label{m8}
\ea

Eqs. (\ref{m3}), (\ref{m4}) and (\ref{m8}) then lead to the tadpole
conditions
\be
\alpha_1 = 32 \quad , \hspace{1cm} \beta_1 = 32 \quad ,
\label{m9}
\ee 
while the corresponding charge parametrization is
\be
\alpha_1 = 2(n_1+n_2)  \quad , \hspace{0.5cm}
\alpha_2 = 2(n_1-n_2)  \quad , \hspace{0.5cm}
\beta_1 = 2(d_1+d_2)  \quad , \hspace{0.5cm}
\beta_2 = 2(d_1-d_2)  \quad .
\label{m10}
\ee
In order to display the spectrum, it is convenient to halve the radius,
reverting to the Scherk-Schwarz 
basis. With the charge assignments (\ref{m10}), the final result for the 
direct-channel annulus amplitude is 
\ba
A \!&=&\! \frac{1}{2}\{ 
[(n_1^2 \!+\! n_2^2) P \!+\! (d_1^2 \!+\! d_2^2) W](Q_O \!+\! Q_V) \!+\! 
2 (n_1d_2 \!+\! n_2d_1)(Q_S \!+\! Q_C) ({Q_S \!+\! Q_C \over 4})_B \} Z_{2m} \nonumber \\ 
\!&+&\! \{ (n_1n_2P \!+\! d_1d_2W)(Q_O \!+\! Q_V) \!+\! (n_1d_1 \!+\! n_2d_2) (Q_S \!+\! Q_C) 
({Q_S \!+\! Q_C \over 4})_B \} Z_{2m+1}  \, , \label{m11}
\ea
while, after a $P$ transformation, the direct-channel M\"obius amplitude is
\ba
M = -\frac{1}{2} \{ 
(n_1 P + d_1 W)(\hat{Q}_O + \hat{Q}_V) -
(n_2+d_2) (\hat{Q}_O - \hat{Q}_V) (\hat{I}_4\hat{I}_4-\hat{V}_4\hat{V}_4)_B \}
Z_{4m}  \nonumber \\
 - \frac{1}{2} \{ (n_2 P + d_2 W)(\hat{Q}_O + \hat{Q}_V) - 
(n_1 +d_1) (\hat{Q}_O - \hat{Q}_V) (\hat{I}_4\hat{I}_4-\hat{V}_4\hat{V}_4)_B 
\} Z_{4m+2} \, .
\label{m12}
\ea

It is instructive to display the massless open spectrum of this model, encoded
in
\ba
&& A = \frac{n_1^2 + n_2^2 + d_1^2 + d_2^2}{2}Q_O 
+ \frac{n_1^2 + n_2^2 + d_1^2 + d_2^2}{2}Q_V
+ (n_1d_2 + n_2d_1)(Q_S + Q_C) \nonumber \\
&& M = - \frac{n_1 - n_2 + d_1 - d_2}{2}\hat{Q}_O 
- \frac{n_1 + n_2 + d_1 + d_2}{2}\hat{Q}_V \quad . \label{m13}
\ea 
It should be appreciated that the gauge group, $SO(n_1) \times Sp(n_2) \times SO(d_1)
\times Sp(d_2)$, with $n_1+n_2=16$, $d_1+d_2=16$, has a reduced rank with respect to
standard orbifold models. Moreover, the first two gauge factors originate
from 16 D9 branes, while the last two originate from 16 D5 branes\footnote{After the 
T-duality to the type I' picture, these become D8 and D4 branes, respectively.}.

There are a number of interesting subtleties in this model. First of all, as
we have seen, the gauge
group has a reduced rank. This phenomenon, familiar in toroidal models \cite{BPS},
where it is induced by quantized expectation values of the NS-NS antisymmetric tensor,
here draws its origin from the asymmetric nature of the winding shifts, that
indeed allows this construction only in the $Z_2$ case. 
Moreover, the open spectrum itself is rather peculiar, since 
away from the origin of the $T^4$ lattice only some of the states are symmetrized in
a conventional fashion.
This is the case, for instance, for the term proportional to $n_1^2 P Z_{4m}$, while
the term proportional to $n_1^2 P Z_{4m+2}$ does not have a corresponding
M\"obius contribution. This unfamiliar projection is actually consistent, since one 
more subtlety helps to settle matters. Namely, the very normalization of the annulus 
is unconventional, as can be seen, for instance, comparing eqs. (\ref{m11}) 
and (\ref{s10}). In toroidal models,
the lattice contributes an operator of the type 
$exp(i p_L x_L + i p_R x_R )$ for each lattice point. 
The conventional $Z_2$ orbifold projection effectively
halves the number of available lattice operators, since the corresponding $cos$ 
and $sin$
terms are assigned opposite eigenvalues. Therefore, in terms of the orbifold 
counting, away from the origin the annulus describes
effectively two copies of the minimal spectrum. The M\"obius
projection on these two copies then realizes the two cases found, for instance, in 
WZW models with
an extended algebra \cite{PSS}: the terms without a M\"obius contribution describe
a pair of sectors, one with symmetric and one with antisymmetric Chan-Paton matrices, 
while the terms with a M\"obius contribution describe a pair of sectors,
both with (anti)symmetric 
Chan-Paton matrices. Finally, the full $N=4$ supersymmetry should be
recovered in the $R \rightarrow 0$ limit, but
in this limit all massive winding modes clearly become massless and give the additional
tadpole conditions $\alpha_1=\alpha_2$, $\beta_1=\beta_2$, or equivalently
$n_2=d_2=0$, in (\ref{m12}) and (\ref{m13}) \cite{ADS,PW}. 
These leave only the N=4 $(Q_O+Q_V)$
character at the massless level, and determine the unique gauge group 
$SO(16) \times SO(16)$, expected by M-theory duality arguments
similar to those presented in \cite{ADS}\footnote{Notice that, in the M-theory model of
\cite{ADS}, the gauge group $SO(16) \times SO(16)$ is also singled out if one
requires that these small-radius singularities be absent.}. 

Letting $n_2=d_2=0$, one can then see 
from (\ref{m11}) that for the $({\bf 16,16})$
mixed representation, corresponding to the term
$n_1d_1 (Q_S+Q_C)(Q_S+Q_C)_B Z_{2m+1}$, the momentum sum is 
shifted by $1/2R$. This Wilson line has a counterpart in the T-dual, type I' model,
that naturally associates the two $SO(16)$ gauge groups
to the two different Horava-Witten walls.
In the M-theory picture related to this type-I' setting, 
partial supersymmetry breaking would then be induced by an R-parity
symmetry coupled to the eleventh dimension ($x_{11}\equiv x_5$). 
Still, by duality arguments one would naively expect 
a Horava-Witten type picture with two D8 branes, each containing an SO(16) gauge group.
In our case, however, the D8 branes on one of the walls are 
actually wrapped around the $T^4$ torus and have effectively
turned into D4 branes (or, before T-duality, into the D5 branes that we have 
found explicitly).

Let us finally mention that in 4d reductions, where
the extra circle $S^1$ is replaced by $T^2$, there are in general several directions 
along $T^2$ that can accommodate the shift, but in the corresponding models the
main physical results remain unchanged.
\section{$N=4 \rightarrow N=1$ Scherk-Schwarz breaking}

One can also break $N=4$ to $N=1$ deforming the compactification
lattice along a pair of different directions. The corresponding operations, however,
must be compatible with the group structure, and the resulting models are rather
involved. Here, we would like to describe
a simpler model that is actually based on a $(T^2)^3$ compactification with shifts along 
three directions \cite{KK}. In this case one has the three group
operations
$g=(\delta_1,-1,-1 \delta_3)$, $h=(-1, -1 \delta_2, \delta_3)$, 
$f=(-1 \delta_1, \delta_2, -1)$ 
with a $Z_2 \times Z_2$ orbifold structure, where $\delta_i$ indicates an order-two 
momentum shift in the compact torus $T^2_i$. 

The amplitudes involve the $16$ characters \cite{MB}
\ba
\tau_{oo}&=&V_2I_2I_2I_2+I_2V_2V_2V_2-S_2S_2S_2S_2-C_2C_2C_2C_2 \ , \nonumber \\
\tau_{og}&=&I_2V_2I_2I_2+V_2I_2V_2V_2-C_2C_2S_2S_2-S_2S_2C_2C_2 \ , \nonumber \\
\tau_{oh}&=&I_2I_2I_2V_2+V_2V_2V_2I_2-C_2S_2S_2C_2-S_2C_2C_2S_2 \ , \nonumber \\
\tau_{of}&=&I_2I_2V_2I_2+V_2V_2I_2V_2-C_2S_2C_2S_2-S_2C_2S_2C_2 \ , \nonumber \\
\tau_{go}&=&V_2I_2S_2C_2+I_2V_2C_2S_2-S_2S_2V_2I_2-C_2C_2I_2V_2 \ , \nonumber \\
\tau_{gg}&=&I_2V_2S_2C_2+V_2I_2C_2S_2-S_2S_2I_2V_2-C_2C_2V_2I_2 \ , \nonumber \\
\tau_{gh}&=&I_2I_2S_2S_2+V_2V_2C_2C_2-C_2S_2V_2V_2-S_2C_2I_2I_2 \ , \nonumber \\
\tau_{gf}&=&I_2I_2C_2C_2+V_2V_2S_2S_2-S_2C_2V_2V_2-C_2S_2I_2I_2 \ , \nonumber \\
\tau_{ho}&=&V_2S_2C_2I_2+I_2C_2S_2V_2-C_2I_2V_2C_2-S_2V_2I_2S_2 \ , \nonumber \\
\tau_{hg}&=&I_2C_2C_2I_2+V_2S_2S_2V_2-C_2I_2I_2S_2-S_2V_2V_2C_2 \ , \nonumber \\
\tau_{hh}&=&I_2S_2C_2V_2+V_2C_2S_2I_2-S_2I_2V_2S_2-C_2V_2I_2C_2 \ , \nonumber \\
\tau_{hf}&=&I_2S_2S_2I_2+V_2C_2C_2V_2-C_2V_2V_2S_2-S_2I_2I_2C_2 \ , \nonumber \\
\tau_{fo}&=&V_2S_2I_2C_2+I_2C_2V_2S_2-S_2V_2S_2I_2-C_2I_2C_2V_2 \ , \nonumber \\
\tau_{fg}&=&I_2C_2I_2C_2+V_2S_2V_2S_2-C_2I_2S_2I_2-S_2V_2C_2V_2 \ , \nonumber \\
\tau_{fh}&=&I_2S_2I_2S_2+V_2C_2V_2C_2-C_2V_2S_2V_2-S_2I_2C_2I_2 \ , \nonumber \\
\tau_{ff}&=&I_2S_2V_2C_2+V_2C_2I_2S_2-C_2V_2C_2I_2-S_2I_2S_2V_2 \ , \label{o1}
\ea
and the torus contribution may be obtained from eq. (\ref{p5}), after 
a further $Z_2$ orbifold projection accompanied by shifts in
the three relevant momenta $(m_1,m_2,m_3)$. The end result is
\ba
T\!\!\!&=&\!\!{1 \over 4} \{ |\tau_{oo}+\tau_{og}+\tau_{oh}+\tau_{of}|^2 \Lambda_1 \Lambda_2 \Lambda_3
+|\tau_{oo}+\tau_{og}-\tau_{oh}-\tau_{of}|^2 (-1)^{m_1} \Lambda_1 
|{4\eta^2 \over \theta_2^2}|^2 \nonumber \\
\!\!\!\!\!\!&+&\!\!\!\!\!\!|\tau_{oo}-\tau_{og}+\tau_{oh}-\tau_{of}|^2 (-1)^{m_3} \Lambda_3 
|{4\eta^2 \over \theta_2^2}|^2+
|\tau_{oo}-\tau_{og}-\tau_{oh}+\tau_{of}|^2 (-1)^{m_2} \Lambda_2 
|{4\eta^2 \over \theta_2^2}|^2 \nonumber \\
\!\!\!\!\!\!&+&\!\!\!\!\!\!|\tau_{go}+\tau_{gg}+\tau_{gh}+\tau_{gf}|^2 \Lambda_1^{n_1+{1 \over 2}}
|{4\eta^2 \over \theta_4^2}|^2+ | \tau_{go}+\tau_{gg}-\tau_{gh}-\tau_{gf}|^2 
(-1)^{m_1} \Lambda_1^{n_1+{1 \over 2}}|{4\eta^2 \over \theta_3^2}|^2 
\nonumber \\
\!\!\!\!\!\!&+&\!\!\!\!\!\!|\tau_{ho}+\tau_{hg}+\tau_{hh}+\tau_{hf}|^2 \Lambda_3^{n_3+{1 \over 2}}
|{4\eta^2 \over \theta_4^2}|^2 +|\tau_{ho}-\tau_{hg}+\tau_{hh}-\tau_{hf}|^2 
(-1)^{m_3} \Lambda_3^{n_3+{1 \over 2}}|{4\eta^2 \over \theta_3^2}|^2  \nonumber \\
\!\!\!\!\!\!&+&\!\!\!\!\!\!|\tau_{fo}\!+\!\tau_{fg}\!+\!\tau_{fh}\!+\!\tau_{ff}|^2 
\Lambda_2^{n_2+{1 \over 2}}|4{\eta^2 \over \theta_4^2}|^2 + 
|\tau_{fo}\!-\!\tau_{fg}\!-\!\tau_{fh}\!+\!\tau_{ff}|^2 
(-1)^{m_2} \Lambda_2^{n_2+{1 \over 2}}|4{\eta^2 \over \theta_3^2}|^2 \} \ , 
\label{o2} 
\ea
where $\Lambda_i$, $(i=1,2,3)$ are the lattice sums for the three 
compact tori and, for instance, the shorthand notation
$(-1)^{m_i} \Lambda_i^{n_i+1/2}$ indicates a sum
with the insertion of $(-1)^{m_i}$ along one of the two momenta of $T^2_i$, 
and with the corresponding windings shifted by $1/2$ unit.
Notice that the torus (\ref{o2}) has indeed the structure of a $Z_2 \times Z_2$ 
orbifold, but all lattice independent terms are 
absent, since the (winding) shifts in the twisted 
sector eliminate all terms at the origin of the lattice for the three tori. 
Equivalently, from a geometric viewpoint, this reflects the nontrivial
action of the projection on the would-be fixed points.
  
The Klein bottle amplitude in the direct channel is then
\be
K\!\!= \!{1 \over 8} (\tau_{oo}+\tau_{og}+\tau_{oh}+\tau_{of}) \{ P_1P_2P_3+
(-1)^{m_1}P_1W_2W_3+W_1(-1)^{m_2}P_2W_3+W_1W_2(-1)^{m_3}P_3 \} \ , \label{o3}
\ee
where $P_1,P_2,P_3$ ($W_1,W_2,W_3$) denote the momentum (winding) sums for the three
tori. By an S transformation, one can obtain the transverse-channel amplitude
\be
\tilde{K}\!\!= \!{2^5 \over 8} (\tau_{oo}+\tau_{og}+\tau_{oh}+\tau_{of}) \{ v_1v_2v_3
W_1^eW_2^eW_3^e+
{v_1 \over v_2v_3}W_1^oP_2^eP_3^e+{v_2 \over v_1v_3}P_1^eW_2^oP_3^e
+ {v_3 \over v_1v_2}P_1^eP_2^eW_3^o \} \, , \label{o4}
\ee
where $v_1,v_2,v_3$ are the volumes of the three compact tori and the superscripts
$e$ ($o$) in (\ref{o4}) indicate the even (odd) parts of the corresponding sums. 
Since the only 
massless tadpole in $\tilde{K}$ is proportional to $v_1v_2v_3$,
the open string spectrum will contain only D9 branes. 
  
>From the torus amplitude, introducing the minimum required number of Chan-Paton 
charges $I_N,g_N,h_N,f_N$, one can deduce the annulus amplitude in the transverse 
channel:
\ba
\tilde{A} \!=\! {2^{-5} \over 8} \{ I_N^2 (\tau_{oo} \!+\! \tau_{og} \!+\! \tau_{oh} \!+\!
\tau_{of})
v_1v_2v_3W_1W_2W_3 \!+\! v_1g_N^2
(\tau_{go} \!+\! \tau_{gg} \!+\! \tau_{gh} \!+\! \tau_{gf})
W_1^{n_1+{1 \over 2}}{4\eta^2 \over \theta_4^2} \nonumber \\
\!+\! v_3h_N^2(\tau_{ho} \!+\! \tau_{hg} \!+\! \tau_{hh} \!+\! \tau_{hf}) 
W_3^{n_3+{1 \over 2}}{4\eta^2 \over \theta_4^2} \!+\!
v_2f_N^2(\tau_{fo} \!+\! \tau_{fg} \!+\! \tau_{fh} \!+\! \tau_{ff}) 
W_2^{n_2+{1 \over 2}}{4\eta^2 \over \theta_4^2} \}
\ . \label{o5}
\ea
The ${4\eta^2 / \theta_4^2}$ factors in (\ref{o5}) originate from four of 
the internal 
bosonic coordinates, actually twisted in three different tori, that we
do not distinguish for brevity while, 
for instance, the superscript in
$W_1^{n_1+1/2}$ indicates that the corresponding winding sum
is shifted in some
direction by $1/2$ unit.  In the same notation,
the direct-channel annulus amplitude then reads (in our convention, in
A and M the momentum sum $P^e$, for instance, actually describes all integer levels)  
\ba
A={1 \over 8} \{ I_N^2 (\tau_{oo}+\tau_{og}+\tau_{oh}+\tau_{of})
P_1^eP_2^eP_3^e +g_N^2(\tau_{oo}+\tau_{og}-\tau_{oh}-\tau_{of}) (-1)^{m_1}P_1^e {4\eta^2 \over \theta_2^2} \nonumber \\
+h_N^2(\tau_{oo}-\tau_{og}+\tau_{oh}-\tau_{of}) (-1)^{m_3}P_3^e {4\eta^2 \over \theta_2^2}+
f_N^2(\tau_{oo}-\tau_{og}-\tau_{oh}+\tau_{of}) (-1)^{m_2}P_2^e {4\eta^2 \over \theta_2^2} \}
\ . \label{o6}
\ea

As in the previous cases, the transverse-channel M\"obius amplitude 
is determined by the characters common to $\tilde{K}$
and $\tilde{A}$. Thus, only the $I_N$ charge contributes to $\tilde{M}$, while
the correct particle
interpretation in the direct channel fixes some sign freedom. The final 
expression is
\ba
\tilde{M}=-{1 \over 4} I_N \{  ({\hat \tau}_{oo}+{\hat \tau}_{og}+{\hat \tau}_{oh}+
{\hat \tau}_{of})
v_1v_2v_3W_1^eW_2^eW_3^e -v_1({\hat \tau}_{oo}+{\hat \tau}_{og}-{\hat \tau}_{oh}-
{\hat \tau}_{of})W_1^o 
{4{\hat \eta}^2 \over {\hat \theta}_2^2}  \nonumber \\
-v_3({\hat \tau}_{oo}-{\hat \tau}_{og}+{\hat \tau}_{oh}-{\hat \tau}_{of})W_3^o 
{4{\hat \eta}^2 \over {\hat \theta}_2^2}-
v_2({\hat \tau}_{oo}-{\hat \tau}_{og}-{\hat \tau}_{oh}+{\hat \tau}_{of}) W_2^o 
{4{\hat \eta}^2 \over {\hat \theta}_2^2} \}
\ , \label{o7}
\ea
and the P-matrix in ref. \cite{BS} determines the direct channel amplitude
\ba
M \!=\! -{1 \over 8} I_N \{  ({\hat \tau}_{oo} \!+\! 
{\hat \tau}_{og} \!+\! {\hat \tau}_{oh}
\!+\! {\hat \tau}_{of})
P_1^eP_2^eP_3^e \!+\! ({\hat \tau}_{oo} \!+\! {\hat \tau}_{og} \!-\! {\hat \tau}_{oh} \!-\! 
{\hat \tau}_{of})
(-1)^{m_1}P_1^e {4{\hat \eta}^2 \over {\hat \theta}_2^2}  \nonumber \\
\!+\! ({\hat \tau}_{oo} \!-\! {\hat \tau}_{og} \!+\! {\hat \tau}_{oh} \!-\! 
{\hat \tau}_{of})(-1)^{m_3}P_3^e 
{4{\hat \eta}^2 \over {\hat \theta}_2^2} \!+\!
({\hat \tau}_{oo} \!-\! {\hat \tau}_{og} \!-\! {\hat \tau}_{oh} \!+\! {\hat \tau}_{of})
(-1)^{m_2}P_2^e {4{\hat \eta}^2 \over {\hat \theta}_2^2} \} \ . 
\label{o8}
\ea

The proper Chan-Paton charge parametrization is in this case
\ba
I_N=n_o+n_g+n_h+n_f \qquad , \qquad g_N= n_o+n_g-n_h-n_f \ , \nonumber \\
h_N= n_o-n_g+n_h-n_f \qquad , \qquad f_N= n_o-n_g-n_h+n_f \ , \label{o9}
\ea
while the tadpole conditions give $I_N=32$, so that the resulting D9 gauge group is
$SO(n_o) \times SO(n_g) \times SO(n_h) \times SO(n_f)$. The massless spectrum
is $N=1$ supersymmetric, and contains, in addition to the vector multiplets,
chiral multiplets in the representations 
$({\bf n_o,n_g,1,1})$, $({\bf 1,1,n_h,n_f})$, $({\bf n_o,1,n_h,1})$, $({\bf 1,n_g,1,n_f})$, 
$({\bf n_o,1,1,n_f})$ and $({\bf 1,n_g,n_h,1})$.
This model can be seen as a discrete deformation of the $N=4$ supersymmetric 
type I model with a Wilson line that breaks the gauge group $SO(32)$ to
$SO(n_o) \times SO(n_g) \times SO(n_h) \times SO(n_f)$. In the particular case 
$n_g=n_h=n_f=0$ one recovers again $SO(32)$, as in the
$N=4 \rightarrow N=2$ example of Section 4.
This model is also dual to a heterotic vacuum
with $N=4$ supersymmetry spontaneously broken to $N=1$ by compactification \cite{KK}, 
as can be seen by arguments similar to those in \cite{ADS}.
\section{Conclusions}

In this paper we have constructed different models with partial and total supersymmetry
breaking in 6d, 5d and 4d type I strings, starting from type IIB
models and deriving by standard methods \cite{S,PS,BS,GP} the corresponding
open descendants. 
In the partial breaking case, the resulting models provide new $N=2$ and $N=1$
supersymmetric type I compactifications, that may also be related by duality arguments
to other types of vacua. In particular, the M-theory 
breaking model
discussed in Section 5 has a natural interpretation in terms
of the eleventh dimension of M-theory \cite{AQ,DG}, while the Scherk-Schwarz breaking 
models
discussed in Sections 4 and 6 are easily interpreted as perturbative breakings in the
dual heterotic language. 

The models present some surprising features. Thus, for
instance,  in the type I' context the
model of Section 5 contains D8 and D4 branes, both having
as origin in the M-theory picture the Horava-Witten 10d boundaries. As another example,
in the $N=4 \rightarrow N=1$ model of Section 6, supersymmetry breaking removes from 
the Klein  bottle the tadpole contributions that in the supersymmetric
$Z_2 \times Z_2$ type I model are related to D5 branes, and the resulting spectrum has 
thus
no D5 branes. More general models contain richer patterns of D5-brane configurations.  
It would be interesting to construct the corresponding effective field theories and to
compare them with the existing field theory models of partial supersymmetry breaking
\cite{APT}.

We have provided additional evidence that the general rule in these 
constructions is that branes with world-volume parallel to the breaking coordinate
feel supersymmetry breaking at tree-level, while branes orthogonal to this coordinate have
a massless spectrum that at tree-level is still supersymmetric. In these sectors, the breaking propagates through
radiative corrections and the resulting picture has some potential applications that are 
interesting for phenomenology, for instance to models with a low string scale, that have
received some attention lately \cite{low}, or to models with the 
``world as a brane'' picture \cite{OS}. It would be interesting to study more 
general 4d examples with chirality \cite{ABPSS,Z6} and $N=1$
supersymmetry spontaneously broken by Scherk-Schwarz compactifications based on
R-symmetries and to investigate whether the chirality flip observed in 
the temperature-like
$(-1)^F$ breaking applies to these cases as well. Finally, a detailed
evaluation of the radiative corrections to some relevant 
quantities in models with massless sectors where supersymmetry is unbroken at
tree level would provide further insight on the
less conventional M-theory breaking mechanism.
\\[5mm]
\noindent{\bf Acknowledgements} 
G.D., E.D. and A.S. are grateful to the Centre de Physique 
Th\'eorique of the Ecole
Polytechnique for the kind hospitality during the course of this work. 
I.A. would like to acknowledge interesting discussions with C. Angelantonj, 
K. Benakli and K. Foerger. Finally, G.D. would like to
thank the Ecole Polytechnique for a CIES Fellowship.

\end{document}